\documentclass[12pt]{article}

\usepackage{amsmath, mathtools}
\usepackage{marvosym}
\usepackage{geometry}
\usepackage{lscape}
\usepackage{afterpage}
\usepackage{amssymb}
\usepackage{multirow}
\usepackage{slashed}
\usepackage{setspace}
\usepackage{fancyhdr}
\usepackage{fourier-orns}
\usepackage[mathcal]{euscript}
\usepackage{indentfirst}
\usepackage{hyperref}
\usepackage{tikz}
\usepackage[all]{hypcap}
\usetikzlibrary{decorations.markings}
\usetikzlibrary{arrows}

\def\cedilla#1{{#1}}    	

\renewcommand{\(}{\left(}
\renewcommand{\)}{\right)}
\newcommand{\spartial}{\slashed{\partial}}
\renewcommand{\bar}[1]{\mkern 3.5mu\overline{\mkern-3.5mu#1\mkern-3.5mu}\mkern 3.5mu}

\newcommand{\D}{D}

\newcommand{\N}{N}

\newcommand{\R}{\mathbb{R}}
\newcommand{\W}{\mathcal{W}}
\renewcommand{\L}{\mathcal{L}}

\renewcommand{\d}{{\rm d}}
\newcommand{\ah}{{\hat{\alpha}}}
\newcommand{\bh}{{\hat{\beta}}}
\newcommand{\gh}{{\hat{\gamma}}}
\renewcommand{\dh}{{\hat{\delta}}}
\newcommand{\xh}{{\hat{\xi}}}

\newcommand{\ha}{{\hat{a}}}
\newcommand{\hb}{{\hat{b}}}
\newcommand{\hc}{{\hat{c}}}
\newcommand{\hd}{{\hat{d}}}
\newcommand{\he}{{\hat{e}}}
\newcommand{\hf}{{\hat{f}}}

\newcommand{\ad}{{\dot{\alpha}}}
\newcommand{\ua}{{\underline{\alpha}}}
\newcommand{\ub}{{\underline{\beta}}}
\newcommand{\ug}{{\underline{\gamma}}}
\newcommand{\ud}{{\underline{\delta}}}
\newcommand{\us}{{\underline{\sigma}}}

\renewcommand{\epsilon}{\varepsilon}
\def\a{\alpha}
\def\b{\beta}

\def\e{\epsilon}

\def\g{\gamma}

\def\s{\sigma}

\def\G{\Gamma}

\def\pa{\partial}      

\def\fracm#1#2{\hbox{\large{${\frac{{#1}}{{#2}}}$}}}

\numberwithin{equation}{section}

\hypersetup{
	colorlinks,
	citecolor = black,
	linkcolor = black,
	urlcolor = black,
	bookmarksopen = true,
	bookmarksopenlevel = \maxdimen
}
\font\ro=cmsy10                          
\def\kcr{{\hbox{\ro \char'170}}}                
\def\ktl{{\hbox{\ro \char'170}}}        
\def\ktr{{\hbox{\ro \char'170}}}        
\def\kbl{{\hbox{\ro \char'170}}}        
\def\kbr{{\hbox{\ro \char'170}}}        
\def\border{                                            
        \setlength{\unitlength}{1mm}
        \newcount\xco
        \newcount\yco
        \xco=-21
        \yco=12
        \begin{picture}(140,0)
        \put(\xco,\yco){$\ktl$}
        \advance\yco by-1
        {\loop
        \put(\xco,\yco){$\kcr$}
        \advance\yco by-2
        \ifnum\yco>-230
        \repeat
        \put(\xco,\yco){$\kbl$}}
        \xco=158
        \yco=12
        \put(\xco,\yco){$\ktr$}
        \advance\yco by-1
        {\loop
        \put(\xco,\yco){$\kcr$}
        \advance\yco by-2
        \ifnum\yco>-230
        \repeat
        \put(\xco,\yco){$\kbr$}}
        \put(-19.5,13){\tiny **University of Maryland * Center for String and
         Particle  Theory* Physics Department***University of Maryland *Center
        for String and Particle  Theory** }
        \put(-19.5,-232.5){\tiny **University of Maryland * Center for String and
         Particle  Theory* Physics Department***University of Maryland *Center
        for String and Particle  Theory** }
        \end{picture}
        \par\vskip-8mm}

\def\headpic{                                           
        \indent
        \setlength{\unitlength}{.4mm}
        \thinlines
        \par
        \begin{picture}(29,16)
        \put(165,16){\line(1,0){4}}
        \put(170,16){\line(1,0){4}}
        \put(180,16){\line(1,0){4}}
        \put(175,0){\line(1,0){4}}
        \put(180,0){\line(1,0){4}}
        \put(185,0){\line(1,0){4}}
        \put(169,0){\line(0,1){16}}
        \put(170,0){\line(0,1){16}}
        \put(179,0){\line(0,1){16}}
        \put(180,0){\line(0,1){16}}
        \put(184,0){\line(0,1){16}}
        \put(185,0){\line(0,1){16}}
        \put(169,16){\oval(8,32)[bl]}
        \put(170,16){\oval(8,32)[br]}
        \put(179,0){\oval(8,32)[tl]}
        \put(185,0){\oval(8,32)[tr]}
        \end{picture}
        \par\vskip-6.5mm
        \thicklines}
\def\endtitle{\end{quotation}\newpage}

\parskip=\medskipamount                 
\def\eqalign#1{\,\vcenter{\openup2\jot \caja
        \ialign{\strut \hfil$\displaystyle{##}$&$
        \displaystyle{{}##}$\hfil\crcr#1\crcr}}\,}
\def\caja{\mathsurround=0pt}
\newcommand{\nd}{{\dot{\nu}}} 
\newcommand{\dd}{{\dot{\delta}}}
\newcommand{\bd}{{\dot{\beta}}} 

\begin{document}

\border\headpic {\hbox to\hsize{\today \hfill
{UMDEPP-014-022}}}
\par \noindent
{ \hfill
}
\par

\begin{center}
{\large\bf Superforms in Five-Dimensional, $\N = 1$ Superspace}\\[.3in]
S.\, James Gates, Jr.\footnote{gatess@wam.umd.edu},
William D. Linch \textsc{iii}\footnote{wdlinch3@gmail.com},
and Stephen Randall\footnote{stephenlrandall@gmail.com}
\\[0.2in]
{\it Center for String and Particle Theory\\
Department of Physics, University of Maryland\\
College Park, MD 20742-4111 USA}
\\[0.8in] 
{\bf ABSTRACT}\\[.01in]
\end{center}
\begin{quotation}
{
We examine the five-dimensional super-de Rham complex with $\N = 1$ supersymmetry. 
The elements of this complex are presented explicitly and related to those 
of the six-dimensional complex in $\N = (1, 0)$ superspace 
through a specific notion of dimensional reduction. 
This reduction also gives rise 
to a second source of five-dimensional super-cocycles that is based on the relative 
cohomology of the two superspaces. 
In the process of investigating these complices, we discover various new features including 
branching and fusion (loops) in the super-de Rham complex,
a natural interpretation of ``Weil triviality", 
$p$-cocycles that are not supersymmetric versions of closed bosonic $p$-forms,
and the opening of a ``gap'' in the complex for $D>4$ in which we find a multiplet of superconformal gauge parameters.
}
\\[.4in]
\thispagestyle{empty}
\endtitle

\clearpage
\pagestyle{fancy}
\fancyfoot[C]{\thepage}
\pagenumbering{arabic}

\section{Introduction}
\setcounter{footnote}{0}

The subject of $p$-forms over superspace manifolds (``super $p$-forms'') had its beginnings in the year of 1977 when a number of 
authors \cite{PFm1,PFm2,PFm3} led by J.\ Wess noted that within the context 
of supergravity and supersymmetric gauge theories, the usual notion of 1-forms could possess extensions in superspace.  
The first two works considered the formal structure and definitions of 
super $p$-forms for only the $p = 1$ case.  There was no guidance 
provided on the extension of the notion of super $p$-forms to $p > 1$.  In that same year, the problem of establishing an integration
theory for super $p$-forms was begun \cite{PFm3}.  In this early, more general discussion 
of super $p$-forms with $p > 1$ there appears to have been little, if any, 
attention paid to the role of constraints.

This situation changed in 1980 when it was shown \cite{pfmSJG} how to construct an entire $\N = 1$ four-dimensional super-de 
Rham complex of super $p$-forms (with $0 < p < 4$) over a supermanifold.  Furthermore, for the first time a set of constraints required for the
irreducibility of the supermultiplets for each value of $p$ was established.  

During this period some authors turned their attention 
to the problem of establishing a theory of integration for super $p$-forms on supermanifolds and significant formal progress was made \cite{GSchwz,BSchwz,Vron,VronFNL}.  
However, in 1997 one of the authors (SJG) put forth the ``Ectoplasmic Integration Theory (EIT)'' 
\cite{ecto,Ectonorcor,Fields} that stressed the role of super $p$-form constraints in integration theory.

The basis for the EIT approach is an assertion about topology.  It is suggested 
that the integration theory over a manifold that realizes supersymmetry must have 
the property that the entire superspace is, at the level of topology, essentially 
indistinguishable from its bosonic submanifold.  This is referred to as ``the ethereal conjecture" and immediately leads to an integration theory that 
necessarily includes elements of cohomology.  As super $p$-forms are inextricably 
linked to cohomological calculations, the EIT approach demands an integration
theory where super $p$-forms play a prominent role.

The EIT approach
is not solely a formal statement of the properties of super $p$-forms and their
theory of integration. In its initial presentations, it was shown to solve a 
problem related to superspace density measures that had been stated by
Zumino.  This was done on the basis of the ethereal conjecture and led
to a superspace analog of Stokes' Theorem, modified appropriately
to hold for both rigid and local supermanifolds.  By now, the 
EIT approach has led to a number of recent practical results that include:
\begin{itemize}
\item[(1)] a highly efficient derivation of supergravity density measures
\cite{SGmeasuRs},

\item[(2)] a superspace formulation for 4D, $\N = 8$ supergravity counterterms
\cite{SGN8m},

\item[(3)] a covariant formulation of 4D, $\N = 4$ supergravity anomalies/divergences \cite{AnomDiVs},

\item[(4)] complete formulations of integration on supermanifolds with boundaries \cite{ectoEdge},

\item[(5)] a supergravity derivation of a minimal unitary representation of the string effective action \cite{ectoREP}, and

\item[(6)] establishing the relationship between superspace integration theory and the picture-changing formalism of superstring theory \cite{ectoPiX}.
\end{itemize}
We believe these all speak powerfully to the motivations behind efforts to understand
as fully as possible the structure of super-de Rham complexes in general. 

We begin this article with a review of superforms in four-dimensional, $N=1$ superspace in section \ref{S:Xspecive}.
In section \ref{sec:5d_forms}, we work out the cocycles of the de Rham complex of five-dimensional, $N=1$ superspace.
This is done sequentially by obstructing the closure conditions on a $p$-cocycle to get a $(p + 1)$-coboundary. 
In the process, we generate the supersymmetric version of closed de Rham $p$-forms for all values of $p$ except for $p = 3$ where we find a 3-cocycle that can be interpreted as a multiplet of superconformal gauge parameters instead. 

In section \ref{sec:dim_red} these cocycles are related to those in the corresponding 
six-dimensional complex \textit{via} dimensional reduction. 
In this reduction, we find a second type of cocycle in the relative cohomology arising from the embedding of the five-dimensional superspace in the six-dimensional one. 
The missing $3$-form can then be interpreted as the 3-cocycle of this relative complex.
Finally, in 
section \ref{sec:5d_content} 
we examine the component fields of the multiplets defined by $p$-form 
field-strengths for $p = 2,\, 3,\, 4$. The 2-form and 4-form are the well-known 
vector and linear multiplets, respectively and are in the super-de Rham 
complex, whereas the $3$-form as found in the relative complex is an on-shell tensor multiplet. 
Our conventions and some useful identities for this superspace are provided in appendix \ref{sec:susy_math}.

\section{A Retrospective \& Prospective Perspective} 
\label{S:Xspecive}
$~~~$ 
There exists a well-known hierarchy of $p$-forms in four-dimensional
spacetime
\vspace{0.01cm}
\begin{center}
\footnotesize
\begin{tabular}{|c|c|}\hline
$~~p ~~$  & $~~p\text{-form} $  \\ \hline
$0$    & $ \varphi$ \\ \hline
$1$   & $ A_{ a}$    \\ \hline
$2$   & $t_{ a   b}$   
\\ \hline
$ 3$  & 
$ {\cal X}_{ a    b   c}$ \\ \hline
$4$  & ${\cal Y}_{ a    b   c   d}$
 \\ \hline  
\end{tabular}
\end{center}
\begin{center}
{Table 1:  4D, $\N = 0$ $p$-form Complex}
\end{center}
where for each value of $p$ there exists a field, respectively denoted 
above by $\varphi$, $A_{ a}$, $t_{ a    b}$,
${\cal X}_{ a    b   c}$, and
${\cal Y}_{ a    b   c   d}$.  Each
such field component is completely antisymmetric on the exchange of
its vector indices and describes a gauge field with field-strength
and gauge transformation
\vspace{0.01cm}
\begin{center}
\footnotesize
\begin{tabular}{|c|c|c|}\hline
$~~p~~$  & Field-Strength  &  ${\rm {Gauge~ Variation~Function }}$  \\ \hline
$0$    & $\pa_{ a} \varphi$ & $c_0$ \\ \hline
$1$   & $  \pa_{ a} A_{ b}  ~-~ 
\pa_{ b} A_{ a}$ & $ \pa_{ a} \lambda $  
\\ \hline
$2$   & $
 \pa_{ a} t_{ b   c}  + \pa_{ b} 
 t_{ c   a}   + \pa_{ c} t_{ a  
  b} $ & $  \pa_{ a} \lambda_{ b}  ~-~ 
\pa_{ b}  \lambda_{ a}$  
\\ \hline
$3$  & $
 \pa_{ a} {\cal X}_{ b   c    d}  
 - \pa_{ b} {\cal X}_{ c   d   a}   
 + \pa_{ c} {\cal X}_{ d   a   b}
  - \pa_{ d} {\cal X}_{ a   b   c}  
$ & $
 \pa_{ a} \lambda_{ b   c}  ~+~ 
 \pa_{ b} \lambda_{ c   a}   ~+~ 
 \pa_{ c} \lambda_{ a   b}
  $ \\ \hline
$4$  & $0$  & 
$
 \pa_{ a} \lambda_{ b   c    d}  
 - \pa_{ b} \lambda_{ c   d   a}   
 + \pa_{ c} \lambda_{ d   a   b}
  - \pa_{ d} \lambda_{ a   b   c}  
 $ \\ \hline  
\end{tabular}
\end{center}
\begin{center}
{Table 2:   4D, $\N = 0$ Field Strengths \& Gauge Variations}
\end{center}
It is seen that all the field-strengths and gauge variations can be
collectively written in the forms

\vspace{0.01cm}
\begin{center}
\footnotesize
\begin{tabular}{|c|c|c|}\hline
Degree & Field-Strength  &  ${\rm {Gauge~ Variation~Function }}$  \\ \hline
$~~p~~$ &
$\tfrac{1}{p!} \pa_{[{ a}_1 |} {\cal P}{}_{| { a}_2 
 \dots   { a}_{p+1} ]}    
$ & $
\tfrac{1}{(p-1)!} \pa_{[{ a}_1 |} \lambda_{| { a}_2 
 \dots   { a}_{p-1} ]}   
 $ \\ \hline  
\end{tabular}
\end{center}
\begin{center}
{Table 3:   4D, $\N = 0$ Field Strengths \& Gauge Variations}
\end{center}
but in the special case of $p$ = 0, the gauge variation is {\em {not}} a
local function.  Instead the quantity $c_0$ is a modulus parameter 
implying the absence of a potential function for the scalar field $\varphi$.  

The results first given in \cite{pfmSJG} established the existence of a complex
among {\em {constrained}} super $p$-form superfields as an extension
of the non-supersymmetric structures above and are summarized 
in the following table.  Super $p$-forms in general possess ``super vector"
indices that take on bosonic and fermionic values as in $A$ = $(a, \a,
{\dot \a})$
\vspace{0.01cm}
\begin{center}
\footnotesize
\begin{tabular}{|c|c|}\hline
$~~p ~~$  & $p$-form Superfield  \\ \hline
$0$    & $ \Gamma $ \\ \hline
$1$   & $ \Gamma_{A}$    \\ \hline
$2$   & $\Gamma_{A  B}$   
\\ \hline
$ 3$  & 
$ \Gamma_{A   B  C}$ \\ \hline
$4$  & ${\Gamma}_{A   B  C  D}$
 \\ \hline  
\end{tabular}
\end{center}
\begin{center}
{Table 4:  4D, $\N = 1$ $p$-form Complex}
\end{center}
where each of the quantities denoted by $\Gamma$ is now a superfield.
In the work of \cite{pfmSJG} a complete listing of all the irreducible 
Lorentz representation for each of the super $p$-forms can be found. 
Each super $p$-form possesses a Bianchi identity, field-strength superfield 
and a corresponding gauge variation that are $\N = 1$ extensions of
the results in Table 3.  These take the forms given in equations
(2.7) through (2.9) of \cite{pfmSJG}.

The major discovery in \cite{pfmSJG} was to identify a
complex of
4D, $\N = 1$ {\em prepotentials}
for the $p$-forms.  These prepotentials had been known
in both super Yang-Mills (the familiar $V$) and supergravity (the familiar 
$H^{a}$) for some time.  
Thus, the result was established that gauge 4D, $\N = 1$ $p$-form superfields 
also have prepotentials and themselves form a
complex without reference to the $p$-forms in Table 4.
\vspace{-0.15cm}
\begin{center}
\footnotesize
\begin{tabular}{|c|c|c|c|}\hline
$~~p ~~$  & ${\rm {Prepotential}}$  & 
${\rm {Field~Strength~SF}}$  &  ${\rm {Gauge~ Variation ~SF}}$  \\ \hline
$0$  &$\Phi$   & $i\fracm12(\Phi-\overline{\Phi})$
 & $c_0$ \\ \hline
$1$  & $V$   & $i\overline{D}^{2}\,D_{\a}\,V $ &
$i\,\fracm12(\Lambda\,-\,\overline{\Lambda})$  \\ \hline
$2$  &$V_{\a}$ & $\fracm12(D^{\a}V_{\a}\,+\,\overline{
D}^{\ad} \overline{V}_{\ad}) $ & $i\,\overline{D}^{2}D_{\a}\,\Lambda$  
\\ \hline
$3$  & $V^{\prime}$ & $D^{2}V'$ & $\fracm12(D^{\a}\Lambda_{\a}
\,+\,\overline{D}^{\ad}\overline{\Lambda}_{\ad})  $ \\ \hline
$4$  & $\Phi^{\prime}$ & $0$  & $D^{2}\Lambda $ \\ \hline  
\end{tabular}
\vspace{0.05in}

{\normalsize Table 5:  4D, $\N = 1$ de Rham Complex}
\end{center}
These prepotentials appear in the geometrical $p$-form superfields
{\it {via}} the following equations

$p$ = 1
$$ \eqalign {
\G_{\a}\,&=\,i \, D_{\a}\, V~~, ~~  V ~=~ {\overline V} ~~, \qquad  \cr
\G_{a}
~&=~\,\fracm14\,\s^{\a\bd}_{a} \left[ \, 
 D_{\a} ~,~ \overline{D}_{\dot \b} \,   \right]   V
~~, } $$

$p$ = 2
$$\eqalign {
\G_{\a\b}\,&=\,  
\G_{\a\bd} \, = \, 0 ~~,\cr
\G_{\a\,b}\,&=\,i\,\s_{b\,\a\dot \gamma}\overline{V}^{\dot \gamma}   ~~,~~~
D_{a}\overline{V}_{\bd}\,=\,0  \cr
\G_{a\,b}\,&=\,i\fracm14\Big[(\s_{a\,b})^{\g\,\delta}D_{\g}V_{\delta}
\,+\,(\bar{\s}_{a\,b})^{{\dot \gamma} \,\dd}\overline{D}\,_{\dot \gamma}\overline{V}
\,_{\dd}\Big] ~~~~.
}$$ 

$p$ = 3
$$\eqalign {
\Gamma_{\alpha \beta \g} \,&=\, \Gamma_{\alpha\beta\,c}\,=\, 
\Gamma_{\alpha \beta \nd}\, \,=\, 0 ~~~,~~~
\cr
\Gamma_{\a\bd\,c}\,&=\,i\,\s_{c\,\a\bd}\,V' ~~,   ~~~    V' ~=~ {\overline V}' ~~, \qquad  \cr
\Gamma_{a\,b\,c}\,&=\,-\,i\,\fracm12(\,\s_{b\,c})_{\a\dot \delta}\,{\overline D}{}^{\dot \delta}V'
~~,\cr
\Gamma_{a\,b\,c}\,&=\,\fracm14\,\e_{a\,b\,c\,d}\,\s^{d\,\b\, {\dot \gamma}}
\left[  \,D_{\b} ~,~ {\overline D}{}_{\dot \gamma}\,  \right] \,V'  
~~,}
$$

$p$ = 4 
$$\eqalign { {~~~~~~~~~}
\G_{\a\b\g\delta}\,&=\,\G_{\a\b {\dot \gamma} \delta }\,=\,\G_{\a\b\g\,d}\,=\,
\G_{\a\b {\dot \gamma} \,d} \,=\,\G_{\a\bd\,c\,d}\,=\, 
D_{\a}  \,\overline{\Phi'}  \,=\,0   ~~~, \cr
\G_{\a\b\,c\,d}\,&=\,i\,\fracm12(\s_{c\, d})_{\a\b}\overline{\Phi'
}~~,\qquad \G_{\b\,d\, e\, f}\,=\,-\,\fracm14\e_{d\, e\, f\, g}\s^{g}_{~\b\dot \g}
\overline{D}\,^{\dot \g}\,\overline{\Phi'}~~,  \cr
\G_{a\, b\, c\, d}\,&=\, i\e_{a\, b\, c\, d}(D^{2}\,\Phi'
\,-\,\overline{D}^{2}\,\overline{\Phi'})~~,}
$$ 
A major unfinished task in supersymmetric field theory is to construct
this complex of prepotentials for all dimensions and all degrees
of extension.

There is a close relation between the 4D, $\N = 2$ and 5D, $\N = 1$ 
superspaces.  Thus, the works of \cite{BisS} and \cite{LinN2} are
closely related to our present considerations.  As the formulation of \cite{BisS}
involves harmonics and as we will not venture in that direction in this
work, we restrict our review to the portion of the work of \cite{LinN2} that
is relevant here.

The work of \cite{LinN2} gave an incomplete presentation of the obstruction
complex.  It explicitly treated the cases of $p$ = 1 and $p$ = 2 and made an
implication for the case of $p$ = 0, but the higher values of $p$ were not treated.
These results are summarized in Table 6.
\vspace{0.01cm}
\begin{center}
\footnotesize
\begin{tabular}{|c|c|c|c|}\hline
$~~p ~~$  & ${\rm {Prepotential}}$  & 
Field-Strength SF  &  ${\rm {Gauge~ Variation ~SF}}$  \\ \hline
$0$  &$\chi{}^{\a \, (i \, j)}{}_k$   & $   D{}_{\a \, k}  \chi{}^{\a \, (j \, k)}{}_i ~+~  
{\overline D}{}_{\dot \a}^k  {\overline 
\chi}{}^{\dot \a \, j}{}_{(i \, k)} $
 & $--$ \\ \hline
$1$  & $V{}_i{}^j$   & $\overline{D}^{(4)}\,D^{(2)}_{i \, j}\, C^{i  k}V_k{}^j $ &
$ D{}_{\a \, k}  \chi{}^{\a \, (j \, k)}{}_i ~+~  {\overline D}{}_{\dot \a}^k  {\overline 
\chi}{}^{\dot \a \, j}{}_{(i \, k)}   $  \\ \hline
$2$  &$\Phi$ & $ i ( \, C^{j k} {D}^{(2)}{}_{i \, k}\, \Phi ~-~
C_{i k} \overline{D}^{(2)}{}^{j \, k}\, {\overline \Phi }) 
 $ & $ \overline{D}^{(4)}\,D^{(2)}_{i \, j}\, C^{i  k}V_k{}^j $  
\\ \hline 
\end{tabular}
\end{center}
\begin{center}
{Table 6:  Known Partial 4D, $\N = 2$ Complex}
\end{center}
Of the superfields that appear in this table there are several points to note.
The superfield $\chi{}^{\a \, (i \, j)}{}_k$ is
a spinorial prepotential that is symmetric on the $i$ and $j$ indices.  At
the time these partial complex result were presented, it was not known 
how to use $\chi{}^{\a \, (i \, j)}{}_k$ to construct a supermultiplet of propagating 
fields.  This is to be contrasted with the case of $\N = 1$ where the 
superfield that appears in the $p = 1$ obstruction superfield transformation
can be used to describe $\N = 1$ supermatter.  However, in the work 
of \cite{MATT} it was shown that such a superfield is capable of describing 
a type of $\N = 2$ hypermultiplet in analogy with superfield
$\N = 1$, $p = 1$ gauge parameter.  The superfield $V_i{}^j $ is
often call the ``Mezin\cedilla{c}escu prepotential'' as it first appeared in the
work of \cite{LMez}.  It is a hermitian traceless matrix on its isospin
indices $i$ and $j$.  Finally, the superfield $\Phi$ in Table 6 is 
chiral ${\overline D}{}_{\dot \a}^i \Phi$ = 0 with respect to 4D, $\N = 2$ supersymmetry.

With the story and background of four-dimensional superforms firmly in mind, we now move towards the complex of forms in five-dimensional, $\N = 1$ superspace. Although the logical conclusion of this line of investigation is the construction of the complex at the level of prepotentials, the first step in the process is the construction of the complex at the level of field-strength superfields. As such, we will content ourselves in this work with the derivation of the constraints on the superfields to which the would-be prepotentials are the unconstrained solutions. Already at this level, we will encounter some unexpected complications and elucidate some features of the five-dimensional super-de Rham complex. As mentioned previously, these include 
branching in the the complex (\S\ref{sec:5d_2form}), 
the existence of a second ``relative cohomology'' complex (\S\ref{sec:rc}),
and even $p$-cocycles that are not the supersymmetrization of $p$-forms (\S\ref{sec:red_mult}). 
As will become apparent, these features are expected to manifest generically in superspaces with $D>4$.

\section{Closed Five-Dimensional Superforms}
\label{sec:5d_forms}
\setcounter{equation}{0}

In this section, we work out the super-de Rham cocycles arising by identifying suitable constraints and obstructing them, starting with the closed 1-form in section \ref{sec:5d_1form}. The components of the $p^\mathrm{th}$ cocycle are related by the superspace Bianchi identities \cite{Gates:1983nr,Buchbinder:1998qv,Wess:1992cp}
\begin{equation}
\label{eq:conv_closure}
	0 = \frac{1}{p!} \D_{[A_1} \omega_{A_2 \ldots A_{p + 1}]} + \frac{1}{2!(p - 1)!} T_{[A_1 A_2 \vert}{}^C \omega_{C \vert A_3 \ldots A_{p + 1}]} .
\end{equation}
This collection is graded by increasing engineering dimension with the component $\omega_{{\underline \alpha}_1 \dots {\underline \alpha}_r a_1 \dots a_s}$, for example, having dimension $\frac r2 + s$. This allows the determination of the higher-dimension components of the cocycle in terms of the lowest non-vanishing one. This lowest non-vanishing component will be a superfield, possibly in a non-trivial (iso-)spin representation. 

In addition to determining the components of the cocycle in terms of this defining superfield, the Bianchi identities generally impose a series of constraints on it, again organized by engineering dimension. As we will see, the highest of these can be obstructed, thereby defining a cocycle of degree 1 higher in the complex.
The complex can branch if it happens that there is more than one constraint on the defining superfield in the highest dimension (as we will see explicitly when passing from the 1-cocycle to the 2-cocycle) and we work out the components of each of the resulting cocycles. 

\subsection{The Five-Dimensional 1-form}
\label{sec:5d_1form}
We begin the construction on the de Rham complex with the 1-form $\omega_A = A_A$. Closure of $A$ is equivalent to the Bianchi identity 
\begin{equation}
\label{eq:A}
	0 = 2 \D_{[A} A_{B]} + T_{AB}{}^C A_{C}.
\end{equation}
The closure condition with the lowest engineering dimension has ${}_{AB} = {}_{\ah i \bh j}$:
\begin{equation}
	0 = \D_{\ah i} A_{\bh j} + \D_{\bh j} A_{\ah i} - 2i \epsilon_{ij} (\Gamma^\ha)_{\ah \bh} A_\ha.
\end{equation}
Since it is symmetric on composite spinor indices, (anti-)symmetrizing on the (iso-)spin indices gives three irreducible parts corresponding to the scalar, anti-symmetric tensor, and vector representations. The first two give the constraints
\begin{equation}
\label{eq:1form_dim1_cons}
	\D^{\ah i} A_{\ah i} = 0 \quad \text{and} \quad \D_{(\ah (i} A_{\bh) j)} = 0,
\end{equation} 
while the third determines the vector component of $A$ in terms of its spinor component
\begin{equation}
	A_\psi = - \frac{i}{8} \D^i \Gamma_\psi A_i.
\end{equation}
If we attempt to partially solve these constraints as $A_{\ah i} = \D_{\ah i} U + \D_\ah^j 
U_{ij}$, then they demand that $\D^2_{\ha \hb} U_{ij} = 0$ and $\D^2_{ij} U^{ij} 
= 0$, respectively, while $U$ remains unconstrained.\footnote{These constraints can be solved in terms of unconstrained prepotentials (cf. {\it e.g.} ref. \cite{LMez}), but we will not need their solution here.
}
The components are then given as
\begin{equation}
\label{eq:1form_comp}
	A_{\ah i} = \D_{\ah i} U + \D_\ah^j 
U_{ij} \quad \text{and} \quad A_\ha = \partial_\ha U - \frac{i}{4} D^2_{\ha ij} U^{ij}.
\end{equation}
The dimension-$\tfrac{3}{2}$ Bianchi identity is solved identically through use of the dimension-1 constraints. The dimension-2 Bianchi identity already holds as well, since
\begin{equation}
	\partial_{[\ha} A_{\hb]} = - \frac{i}{4} \partial_{[\ha} D^2_{\hb] ij} U^{ij} = \frac{1}{16} [D^2_{ij}, D^2_{\ha \hb}] U^{ij} = 0.
\end{equation}
Thus, the components \eqref{eq:1form_comp} and constraints \eqref{eq:1form_dim1_cons} together give a closed 1-form field-strength in five dimensions.

\subsection{The Five-Dimensional 2-form}
\label{sec:5d_2form}

The closed 2-form $F = \d A$ is the exterior derivative of a 
gauge 1-form $A$ and can be interpreted, therefore, as the obstruction to the 1-form's closure. By 
setting the lowest component of $F$ to be the obstruction to the scalar constaint in \eqref{eq:1form_dim1_cons}, we have
\begin{equation}
	F_{\ah i \bh j} = (\d A)_{\ah i \bh j} =: 2i \epsilon_{ij} \epsilon_{\ah \bh} \W ,
\end{equation}
for some dimension-1 field-strength $\W$. Now that we have the lowest component of $F$, the remaining components and any constraints on $\W$ follow uniquely from \eqref{eq:conv_closure}. 
For purposes of exposition, we will give a fairly in-depth look at the calculations that go into this analysis in this section, but we will suppress the analogous steps in the following sections. 

To begin, consider the dimension-$\tfrac{3}{2}$ condition
\begin{equation}
	0 = \D_{\ah i} F_{\bh j \gh k} + 2i \epsilon_{ij} (\Gamma^\ha)_{\ah \bh} F_{\gh k \ha} + (\underline{\alpha \beta \gamma}) .
\end{equation}
Here $\underline{\alpha} \equiv \ah i$ and the notation $(\underline{\,\cdot\,})$ denotes the remaining cyclic permutations of the enclosed composite indices. Plugging in $F_{\ah i \bh j}$, we find that $F_{\ah i \ha}$ is fixed to be
\begin{equation}
	F_{\ah i \ha} = - (\Gamma_\ha)_\ah{}^\bh \D_{\bh i} \W .
\end{equation}
The dimension-2 condition, upon plugging in the known components and expanding the $D D$ terms with \eqref{eq:dd_exp}, becomes
\begin{align}
\label{eq:2form_van_dim2_exp}
	0 & = [ - i \epsilon_{ij} (\Gamma_\ha \Gamma^\hb)_{\bh \ah} \partial_\hb - \frac{1}{2} \epsilon_{ij} (\Gamma_\ha \Sigma^{\hb \hc})_{\bh \ah} \D^2_{\hb \hc} + \frac{1}{2} (\Gamma_\ha \Gamma^\hb)_{\bh \ah} \D^2_{\hb i j} - \frac{1}{2} (\Gamma_\ha)_{\bh \ah} \D^2_{ij} \notag\\[4pt]
		& \quad +\, (\underline{\alpha \beta}) ] \W - 2 i \epsilon_{ij} \epsilon_{\ah \bh} \partial_\ha \W + 2i \epsilon_{ij} (\Gamma^\hb)_{\ah \bh} F_{\hb \ha} .
\end{align}
The $(\underline{\alpha \beta})$ symmetry kills the final term in the $\D \D$ expansion and allows the $\partial \W$ terms to cancel. Additionally, it restricts the irreducibles in the remaining two terms of the $\D \D$ expansion, leaving behind the relation
\begin{equation}
	0 = [- \epsilon_{ij} (\Gamma^\hb)_{\bh \ah} \D^2_{\ha \hb} - 2 (\Sigma_\ha{}^\hb)_{\ah \bh} \D^2_{\hb ij}] \W + 2i \epsilon_{ij} (\Gamma^\hb)_{\ah \bh} F_{\hb \ha} .
\end{equation}
Because of the (anti-)symmetry in the $ij$ indices, this is actually two separate conditions with one defining the component $F_{\ha \hb}$ and the other putting a restriction on $\W$. The former yields
\begin{equation}
	F_{\ha \hb} = - \frac{i}{2} \D^2_{\ha \hb} \W ,
\end{equation}
while the latter requires
\begin{equation}
\label{eq:2form_dim2_acons}
	\D^2_{\ha ij} \W = 0 .
\end{equation}
From \eqref{eq:dd_exp}, this is equivalent to
\begin{equation}
\label{eq:2form_avm_cons}
	\D_\ah^{(i} \D_\bh^{j)} \W = \frac{1}{4} \epsilon_{\ah \bh} \D^{\gh (i} \D_\gh^{j)} \W .
\end{equation}
Continuing with the dimension-$\tfrac{5}{2}$ condition, we substitute the components of $F$ to find
\begin{equation}
\label{eq:2form_dim52_bi}
	\D_{\ah i} \D^k_{(\bh} \D_{\gh) k} \W = 4i \spartial_{\dh (\bh} \epsilon_{\gh) \ah} \D^\dh_i \W - 4i \spartial_{\ah (\bh} \D_{\gh) i} \W .
\end{equation}
Through a bit of $\Gamma$-matrix algebra this can be shown to come directly from \eqref{eq:2form_dim2_acons} by expanding and simplifying
\begin{equation}
\label{eq:2form_dim52_check}
	(\Gamma_\ha)_{\ah \bh} (\Gamma_\hb)_{\gh \dh} (\Sigma^{\ha \hb})_{\hat{\rho} \hat{\tau}} \D^{\bh i} \D^\gh_{(i} \D^\dh_{j)} \W = 0 .
\end{equation}
The dimension-3 closure condition, like the dimension-$\tfrac{5}{2}$ condition \eqref{eq:2form_dim52_bi}, holds identically since
\begin{equation}
\label{eq:VMBI3}
	\epsilon_{\ha \hb}{}^{\hc \hd \he} \partial_\hc F_{\hd \he} = - \frac{i}{2} \epsilon_{\ha \hb}{}^{\hc \hd \he} \partial_\hc \D^2_{\hd \he} \W = \frac{1}{12} [\D^2_{\ha ij}, \D^{2 ij}_\hb] \W = 0 .
\end{equation}
Thus, the only constraint on $\W$ is \eqref{eq:2form_dim2_acons} which, as we review in section \ref{sec:vect_mult}, identifies it as the field-strength of the off-shell vector multiplet in five dimensions.

\subsubsection{An Alternative 2-Cocycle}
\label{sec:5d_2form_alt}

Instead of obstructing the first constraint in (\ref{eq:1form_dim1_cons}),
we may define
\begin{equation}
	\tilde{F}_{\ua \ub} = (\Sigma^{\ha \hb})_{\ah \bh} C_{\ha \hb ij} 
\end{equation}
and proceed with this as our lowest 
component. 
Repeating the previous analysis,
the remaining components are found to be
\begin{equation}
\label{eq:a2form_comp}
	\tilde{F}_{\ua \ha} = \frac{i}{12} \epsilon_\psi{}^{\ha \hb \hc \hd} (\Sigma_{\ha 
	\hb})_\ah{}^\bh \D_\bh^j C_{\hc \hd ij} \quad \text{and} \quad \tilde{F}_{\ha \hb} = - \frac{1}{48} \epsilon_{\ha \hb}{}^{\hc \hd \he} \D^2_{\hc 
	ij} C^{ij}_{\hd \he} .
\end{equation}
The dimension-1 field-strength $C_{\ha \hb ij}$ is constrained by the dimension-$\tfrac{3}{2}$ Bianchi identity to satisfy
\begin{equation}
\label{eq:a2form_cons1}
	(\Sigma_{\ha \hb})_{(\ah \bh} \D_{\gh)(i} C^{\ha \hb}_{jk)} = 0
\end{equation}
and by the dimension-2 Bianchi identity to satisfy
\begin{equation}
\label{eq:a2form_cons2}
	6i \partial^\hb C_{\ha \hb ij} + \D^{2 \hb k}_{(i} C_{j) k \ha \hb} - 2 \D^2_{\ha \hb
	 \hc} C^{\hb \hc}_{ij} = 0 .
\end{equation}
The first of these, \eqref{eq:a2form_cons1}, can be re-cast in the form
\begin{equation}
\label{eq:2form_cons2}
	\Pi_{\ha \hb \ah}^{~ \hc \hd \bh} D_{\bh (i} C_{\hc \hd jk)} = 0,
\end{equation}
where
\begin{equation}
\label{eq:pi_stls}
	\Pi_{\ha \hb \ah}^{~ \hc \hd \bh} := \delta_{[\ha}^\hc \delta_{\hb]}^\hd \delta_\ah^\bh + \frac{1}{5} (\Sigma_{\ha \hb} \Sigma^{\hc \hd})_\ah{}^\bh
\end{equation}
is the projection operator onto the $\Sigma$-traceless subspace of the ($2$-form)$\otimes$(spinor) representation space. 
With these constraints in place, the top two Bianchi identities (at dimensions $\tfrac{5}{2}$ and 3) do not imply any new conditions on $C_{\ha \hb ij}$.

\subsection{The Five-Dimensional 3-cocycle}
\label{sec:5d_3form1}
We have obstructed the closure of the 1-form potential in two independent ways and found that each of these is obstructed in turn. The new constraints \eqref{eq:2form_dim2_acons} and \eqref{eq:a2form_cons2} are both dimension-2, vector-valued, isotriplet superfields. To generate the 3-form, we obstruct the closure of the 2-form as $H = \d F$ in either incarnation. The 
components of $H$ are then uniquely 
determined to be
\begin{align}
\begin{array}{lcl}
	H_{\ua \ub \ug} = 0 , && \quad H_{\ua \ub \ha} = (\Sigma_{\ha \hb})_{\ah \bh} H_{ij}^\hb, \\
	H_{\ua \ha \hb} = \frac{i}{12} \epsilon_{\ha \hb}{}^{\hc \hd \he} (\Sigma_{
	\hc \hd})_\ah{}^\bh \D_\bh^j H_{\he ij} , && \quad H_{\ha \hb \hc} = \frac{1}{48} \epsilon_{\ha \hb \hc}{}^{\hd \he} \D_{\hd 
	i j}^2 H_\he^{ij} ,
\end{array}	
\end{align}
where the dimension-2 field $H_{\ha ij}$ satisfies the conditions
\begin{equation}
\label{eq:3form_cons1}
	(\Sigma_{\ha \hb})_{(\ah \bh} \D_{\gh) (i} H_{jk)}^\hb = 0
\end{equation}
at dimension $\tfrac{5}{2}$ and
\begin{equation}
\label{eq:3form_cons2}
	\D^2_{\ha k (i} H_{j)}^{\ha k} + 6 i \partial_\ha H^\ha_{ij} = 0
\end{equation}
at dimension 3.

The way in which the constraints ``fit together" here is fairly interesting. At dimension $\tfrac{5}{2}$, it is not difficult to see that \eqref{eq:3form_cons1} is equivalent to
\begin{equation}
\label{eq:3form_cons3}
	\Pi_{\ha \ah}^{~ \hb \bh} \D_{\bh (i} H_{\hb jk)} = 0 ,
\end{equation}
where
\begin{equation}
\label{eq:pi_gtls}
	\Pi_{\ha \ah}^{~ \hb \bh} ~:=~ \delta_\ha^\hb \delta_\ah^\bh + \frac{1}{5} (\Gamma_\ha \Gamma^\hb)_\ah{}^\bh
\end{equation}
is the projection operator, this time onto the $\Gamma$-traceless subspace of the (vector)$\otimes$(spinor) representation. 
The question, then, is: What part of the dimension-3 Bianchi identity does this already imply, and what part is an independent constraint? If we look at the dimension-3 closure condition more carefully, we find three independent conditions: equation \eqref{eq:3form_cons2} and the following two ``constraints"
\begin{align}
\label{eq:3form_dim3_symm}
	0 & = \D^2_{(\ha k (i} H_{\hb) j)}^{~k} - 4 i \partial_{(\ha} H_{\hb) ij} - \text{trace} , \\[4pt]
\label{eq:3form_dim3_as}
	0 & = \D^2_{[\ha k (i} H_{\hb] j)}^{~k} - 4 i \partial_{[\ha} H_{\hb] ij} - \frac{1}{6} \epsilon_{\ha \hb \hc \hd \he} \D^{\xh k} (\Sigma^{\hc \hd})_\xh{}^\gh \D_{\gh k} H_{ij}^\he .
\end{align}
However, these two conditions follow from \eqref{eq:3form_cons3} in the form 
\begin{equation}
	\D_\xh^k (\Gamma_\hc)^{\xh \ah} \Pi_{\ha \ah}^{~ \hb \gh} \D_{\gh (k} H_{\hb ij)} = 0
\end{equation}
by taking the appropriate index (anti-)symmetrizations. Since the $\Pi$-projector only spits out parts that are symmetric-traceless and anti-symmetric, it leaves \eqref{eq:3form_cons2} untouched and we find it as an independent constraint at dimension 3.

\subsection{The Five-Dimensional 4- and 5-forms}
\label{sec:5d_45forms}

Having found that the constraint \eqref{eq:3form_cons2} on the 3-form at dimension 3 is independent of the lower-dimensional conditions \eqref{eq:3form_cons1}, we can obstruct the closure of that form by introducing a Lorentz-singlet, iso-spin triplet superfield $G_{ij}$ of dimension $3$.
In terms this superfield, the closed 4-form $G$ has components
\begin{align}
\begin{array}{ll}
	G_{\ua \ub \ug \ud} = 0 , & G_{\ua \ub \ug \ha} = 0 , \\
	G_{\ua \ub \ha \hb} = (\Sigma_{\ha \hb})_{\ah \bh} G_{ij} , & G_{\ua \ha \hb \hc} = \frac{i}{12} \epsilon_{\ha \hb \hc}{}^{\hd \he} (
	\Sigma_{\hd \he})_\ah{}^\bh \D_\bh^j G_{ij} , \\
	G_{\ha \hb \hc \hd} = - \frac{1}{48} \epsilon_{\ha \hb \hc \hd}{}^\he 
	\D_{\he ij}^2 G^{ij} , &
\end{array}
\end{align}
in agreement with reference \cite{ectoEdge}. At dimension $\tfrac{7}{2}$, the condition
\begin{equation}
\label{eq:4form_cons}
	\D_{\ah (i} G_{jk)} = 0
\end{equation}
is imposed. All remaining Bianchi identities are then satisfied, with the dimension-5 condition coming from
\begin{equation}
	\partial^\ha (\star G)_\ha = \partial^\ha \D^2_{\ha ij} G^{ij} = \frac{3i}{16} \D^{3 \ah}_{ijk} \D_\ah^k G^{ij} = 0 ,
\end{equation}
where $\star G$ stands for the bosonic Hodge dual of the 4-form components $G_{\hat a \hat b \hat c \hat d}$.

To complete the complex, we proceed in the established way by obstructing the 4-form's defining 
condition as $K = \d G$. Note that this is slightly different than the previous 
obstructions since now the lowest component $K$ stays at the same level as 
that of $G$. This is required for the lowest Bianchi identity to be non-trivially 
satisfied. We then have a closed 5-form $K$ with components
\begin{align}
\begin{array}{l}
	K_{\ua \ub \ug \ud \us} = 0 , \qquad K_{\ua \ub \ug \ud \ha} = 0 , \qquad K_{\ua \ub \ug \ha \hb} = (\Sigma_{\ha \hb})_{\ah \bh} K_{\gh ijk} , \\
	K_{\ua \ub \ha \hb \hc} = - \frac{i}{48} \epsilon_{\ha \hb \hc}{}^{\hd 
	\he} (\Sigma_{\hd \he})_\ah{}^\gh (3 \D_\gh^k K_{\bh ijk} - \D_\bh^k 
	K_{\gh ijk}) , \\
	K_{\ua \ha \hb \hc \hd} = - \frac{1}{192} \epsilon_{\ha \hb \hc \hd}{
	}^\he (2 \D_\he^{2 jk} K_{\ah ijk} + (\Sigma_{\he \hf})_\ah{}^\bh \D^{2 
	\hf jk} K_{\bh ijk}) , \\
	K_{\ha \hb \hc \hd \he} = \frac{i}{768} \epsilon_{\ha \hb \hc \hd \he} 
	\D^3_{\ah ijk} K^{\ah ijk} ,
\end{array}
\end{align}
where the dimension-$\tfrac{5}{2}$ field $K_{\ah ijk}$ satisfies the condition
\begin{equation}
\label{eq:5form_cons}
	\D_{(\ah (i} K_{\bh) jkl)} = 0
\end{equation}
through which all the other Bianchi identities are satisfied.

With this, we have found the structure of all the cocycles in super-de Rham complex of the five-dimensional, $N=1$ superspace. In the process, we found that the sequence splits, giving rise to two 2-cocycles due to the existence of two independent constraints (\ref{eq:1form_dim1_cons}) on the components of the 1-cocycle. These 2-cocycles each have a constraint  on their components at dimension 2 that that are isomorphic as superfield representations: Both equations (\ref{eq:2form_dim2_acons}) and (\ref{eq:a2form_cons2}) are iso-spin triplets of vectors. Because of this, the 3-cocycle resulting from obstructing these equations is unique and the branching fuses. Its dimension-3 constraint (\ref{eq:3form_cons2}) is unique as a superfield representation and can be sourced to uniquely define the iso-spin triplet field-strength of the 4-cocycle. This uniqueness persists to the 5-cocycle. We summarize this structure of the five-dimensional, $N=1$ super-de Rham complex in figure \ref{fig:branch5D}.

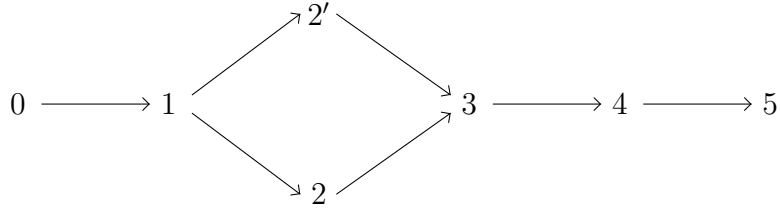
\begin{figure}[h]
\begin{center}
\begin{tikzpicture}[auto, > = {angle 90}, decoration = {
    markings,
    mark = at position 1 with {\arrow{>}}}, scale = 0.8]
    
    \draw (-2.5, 0) node[right] {$0$};
    \draw (0, 0) node[right] {$1$};
    \draw (2.5, 1.5) node[right] {$\hspace{-0.5mm}2^\prime$};
    \draw (2.5, -1.5) node[right] {$2$};
    \draw (5, 0) node[right] {$3$};
    \draw (7.5, 0) node[right] {$4$};
    \draw (10, 0) node[right] {$5$};
	
	\draw [postaction = {decorate}] (-1.8, 0) -- (-0, 0);
	\draw [postaction = {decorate}] (0.7, 0.15) -- (2.5, 1.5);
	\draw [postaction = {decorate}] (0.7, -0.15) -- (2.5, -1.5);
	\draw [postaction = {decorate}] (3.1, 1.5) -- (5, 0.15);
	\draw [postaction = {decorate}] (3.1, -1.5) -- (5, -0.15);
	\draw [postaction = {decorate}] (5.7, 0) -- (7.5, 0);
	\draw [postaction = {decorate}] (8.2, 0) -- (10, 0);
\end{tikzpicture}
\vspace{0.5cm}\\

\caption{The general ``obstruction structure" of the five-dimensional super-de Rham complex as constructed in this article.
\label{fig:branch5D}}
\end{center}
\end{figure}

\section{Dimensional Reduction}
\label{sec:dim_red}

For the computation of the 4- and 5-forms in the previous section, an alternative to the usual procedure was employed that allowed us to determine the components and constraints on the forms by reducing them from a higher-dimensional complex. The observation is that the five-dimensional, $N=1$ de Rham complex 
has a simple interpretation as a specific part of the dimensional reduction of 
of the six-dimensional, $N=(1,0)$ de Rham complex studied in \cite{6dforms}. 
To see this, consider the generic form of a Bianchi identity for a closed
$p$-form $\omega$ in flat 6D superspace. This identity is formally identical to \eqref{eq:conv_closure} as this formula makes no explicit reference to the dimension. Written in 5 + 1 dimensions this splits into two equations:
\begin{align}
\label{eq:red_bi_vec}
	0 & = \frac{1}{p!} \D_{[A_1} \omega_{A_2 \ldots A_{p + 1}]} + \frac{1}{2!(p - 1)!} T_{[A_1 A_2 \vert}{}^C \omega_{C \vert A_3 \ldots A_{p + 1}]} \notag\\
	& \quad + \frac{1}{2!(p - 1)!} T_{[A_1 A_2 \vert}{}^6 \omega_{6 \vert A_3 \ldots A_{p + 1}]}, \\[8pt]
\label{eq:red_bi_6}
	0 & = \frac{1}{p!} \partial_6 \omega_{[A_1 \ldots A_p]} - \frac{1}{(p - 1)!} \D_{[A_1 \vert} \omega_{6 \vert A_2 \ldots A_p]} + \frac{1}{(p - 1)!} T_{6 [A_1 \vert}{}^C \omega_{C \vert A_2 \ldots A_p]} \notag\\
	& \quad - \frac{1}{2! (p - 2)!} T_{[A_1 A_2 \vert}{}^C \omega_{6 C \vert A_3 \ldots A_p]} .
\end{align}
Restricting the vector indices to five dimensions and setting $\partial_6$ and $T_{6 A}{}^B$ to zero suggests the following definitions: the five-dimensional $p$-form
\begin{equation}
	(\alpha_p)_{A_1 \ldots A_p} := \omega_{A_1 \ldots A_p}
\end{equation}
and the five-dimensional $(p - 1)$-form
\begin{equation}
	(\beta_{p - 1})_{A_1 \ldots A_{p - 1}} = \omega_{6 A_1 \ldots A_{p - 1}} .
\end{equation}
The $(5 + 1)$-dimensional closure conditions then give, in an index-free notation,
\begin{equation}
\label{eq:dim_red_rel_coh}
	\d \alpha_p = c_2 \wedge \beta_{p - 1}
~~~\mathrm{and}~~~	
	\d \beta_{p - 1} = 0 ,
\end{equation}
where $c_{\ua \ub} = T_{\ua \ub}{}^6 = \epsilon_{ij} \epsilon_{\ah \bh}$ is the only non-zero component of the constant 2-form $c_2$.

The first thing to notice here is that although two forms come from this reduction, only $\beta_{p - 1}$ 
is closed. Looking back to the complex worked out in section \ref{sec:5d_forms}, the $\beta_{p - 1}$ forms---as they came from six dimensions---are precisely those forms that we studied in section \ref{sec:5d_forms}. For ease of comparison, we have collected the schematic form of the five- and six-dimensional cocycles in the table on the following page. For clarity of presentation, we have suppressed real numerical factors and are using $\star$ to schematically denote factors of $\epsilon_{a_1 \dots a_D}$. The precise forms of the $\Pi$-projectors are given in \eqref{eq:pi_stls} and \eqref{eq:pi_gtls} for five dimensions and in \cite{6dforms} for six. 

\afterpage{
\includegraphics[scale=1.0, trim=3.8cm 0 0 2.5cm]{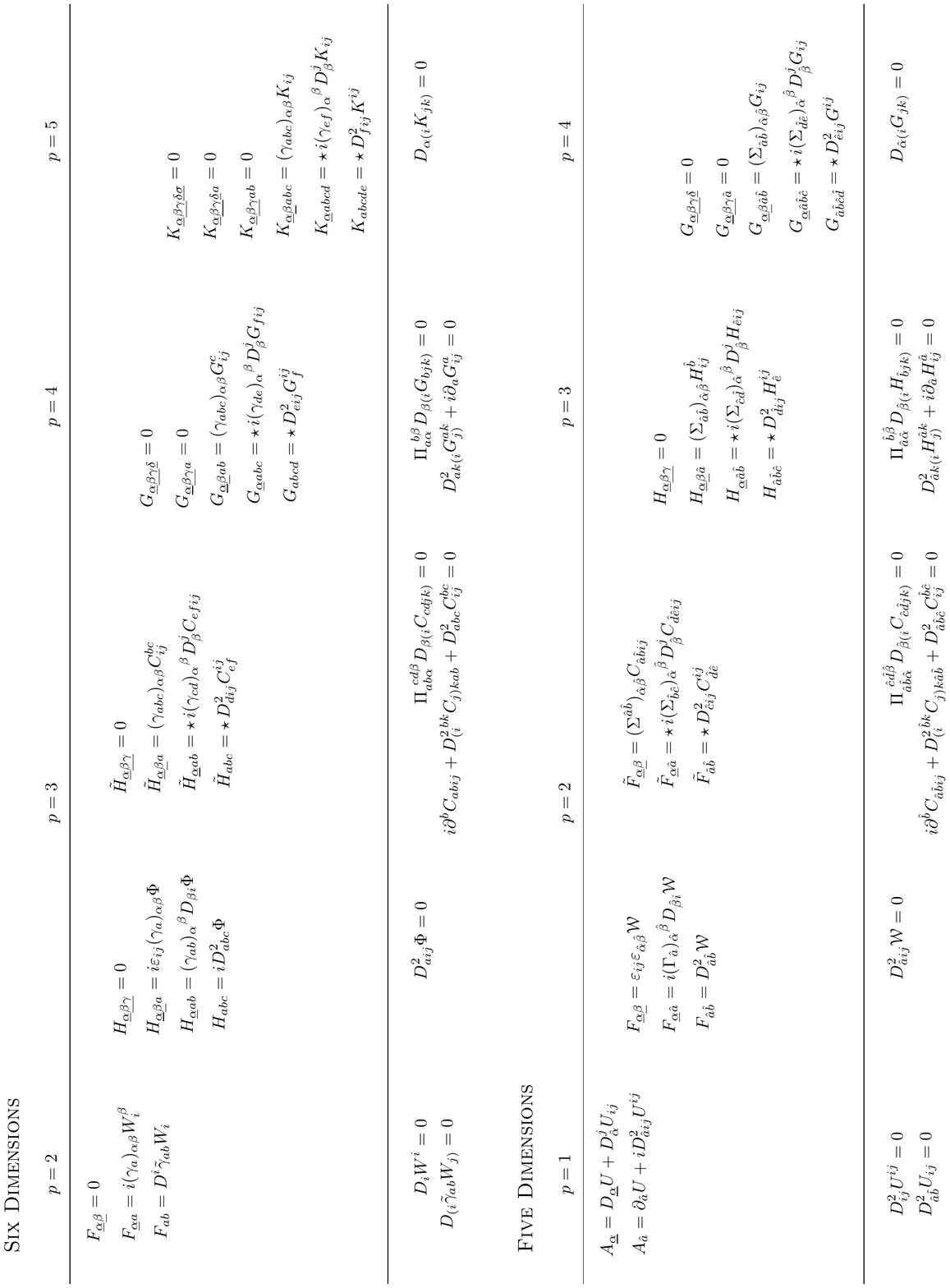}
\thispagestyle{empty}}

Note that the branching structure of the five-dimensional de Rham complex represented by figure \ref{fig:branch5D} descends from a similar branching in the six-dimensional complex (cf. figure \ref{fig:branching6D}) where there are two irreducible constraints for the closed 2-form.\footnote{The second 3-form presented in the table appeared only as a composite 3-form in reference \cite{6dforms}.}
\vspace{0.3cm}

\begin{figure}[htb]
\begin{center}
\begin{tikzpicture}[auto, > = {angle 90}, decoration = {
    markings,
    mark = at position 1 with {\arrow{>}}}, scale = 0.76]
    
    \draw (-5, 0) node[right] {$0$};
    \draw (-2.5, 0) node[right] {$1$};
    \draw (0, 0) node[right] {$2$};
    \draw (2.5, 1.5) node[right] {$3^\prime$};
    \draw (2.5, -1.5) node[right] {$3$};
    \draw (5, 0) node[right] {$4$};
    \draw (7.5, 0) node[right] {$5$};
    \draw (10, 0) node[right] {$6$};
	
	\draw [postaction = {decorate}] (-4.3, 0) -- (-2.5, 0);
	\draw [postaction = {decorate}] (-1.8, 0) -- (0, 0);
	\draw [postaction = {decorate}] (0.7, 0.15) -- (2.5, 1.5);
	\draw [postaction = {decorate}] (0.7, -0.15) -- (2.5, -1.5);
	\draw [postaction = {decorate}] (3.2, 1.5) -- (5, 0.15);
	\draw [postaction = {decorate}] (3.2, -1.5) -- (5, -0.15);
	\draw [postaction = {decorate}] (5.7, 0) -- (7.5, 0);
	\draw [postaction = {decorate}] (8.2, 0) -- (10, 0);
\end{tikzpicture}
\vspace{0.5cm}

\caption{The general ``obstruction structure" of the six-dimensional super-de Rham complex as constructed in reference \cite{6dforms}.
\label{fig:branching6D}}
\end{center}
\end{figure}
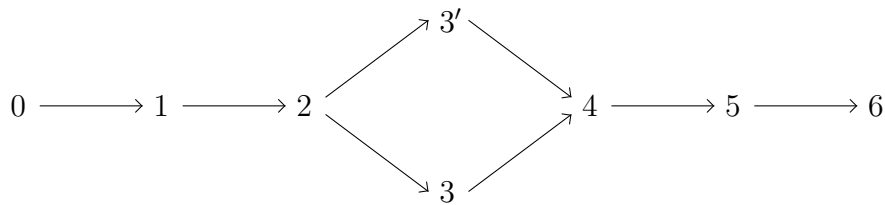
\vspace{-0.3cm}

\subsection{Relative Cohomology}
\label{sec:rc}
Returning to the remaining equation in the reduction (\ref{eq:dim_red_rel_coh}), we 
note that it is possible to 
construct another closed 5D $p$-form by 
solving the closure condition $\d \beta_{p - 1} = 0$ as $\beta_{p - 1} = \d \theta_{p - 2}$ and using this to
define the shifted superform
\begin{equation}
\label{eq:RelativeForm}
	\alpha'_p ~:=~ \alpha_p - c_2 \wedge \theta_{p - 2} .
\end{equation}
The structure of these forms is illustrated in figure \ref{fig:rel_coh}.

Interestingly, we recognize this as
the form that comes from the relative cohomology 
construction of a closed 5-form in reference \cite{ectoEdge}. The fact that their $L_6 = 
c_2 \wedge G_4$ exhibits Weil triviality as $L_6 = \d K_5$ and $L_6 = 
c_2 \wedge \d h_3$ is then a direct consequence of the fact that $G_4$ and $K_5$ come to 5D together as a relative cohomology pair from the dimensional reduction of the 6D 5-form.

\begin{figure}[t]
\begin{center}
\hspace*{0.85cm}
\begin{tikzpicture}[rotate = 90, scale = 2]
	\draw [color = black!50!white] (-0.65, 2) -- (0.65, 2);
	\draw [color = black!50!white] (-0.65, 1.2) -- (0.65, 1.2);

	\draw (-0.5, 3) -- (-0.5, 1.85);
	\draw (-0.505, 1.785) node[right] {$\cdots$};
	\draw (-0.5, 1.35) -- (-0.5, 0.55);
	\draw (-0.505, 0.485) node[right] {$\cdots$};
	
	\draw (0.5, 3) -- (0.5, 1.85);
	\draw (0.495, 1.785) node[right] {$\cdots$};
	\draw (0.5, 1.35) -- (0.5, 0.55);
	\draw (0.495, 0.485) node[right] {$\cdots$};
	
	\draw [fill = black] (-0.5, 3) circle (0.05);
	\draw [fill = black] (-0.5, 2.5) circle (0.05);
	\draw [fill = white] (-0.5, 0.7) circle (0.05);
	\draw (-0.52, -0.125) node[right] {$\alpha_p$};
	\draw (0.52, -0.125) node[right] {$c_2 \wedge \theta_{p - 2}$};
	
	\draw [fill = white] (0.5, 3) circle (0.05);
	\draw [fill = white] (0.5, 2.5) circle (0.05);
	\draw [fill = black] (0.5, 0.7) circle (0.05);
	
	\draw [fill = black] (-0.5, 2) circle (0.05);
	\draw [fill = black] (-0.5, 1.2) circle (0.05);
	\draw [fill = black] (0.5, 2) circle (0.05);
	\draw [fill = black] (0.5, 1.2) circle (0.05);
\end{tikzpicture}
\end{center}
\caption{Filled nodes are the non-zero components of the indicated 
forms, with the struts indicating which components of the $\alpha_p$ are 
``corrected" by $c_2 \wedge \theta_{p - 2}$ to allow the form $\alpha'_p$ to close 
without vanishing. Higher-dimensional components are on the left.
\label{fig:rel_coh}}
\end{figure}
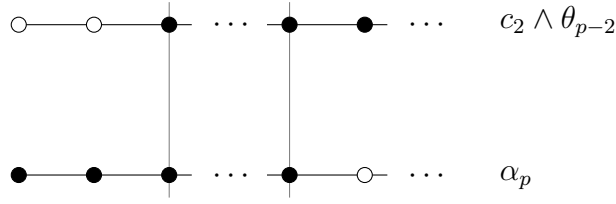

To illustrate this relative cohomology construction and its origin from dimensional reduction, consider the case of the relative 3-form.
It is obtained by reducing the 
six-dimensional 3-form $H\to (H, F)$ to a five-dimensional 3-form $H$ and 2-form $F$. The resulting closed 2-form $F$ is solved in terms of its potential $A$, which is used to correct the non-closed part $H$ of the 3-form as expressed by equation (\ref{eq:RelativeForm}). 
The closed 3-form $H'$ arising from this construction has components
\begin{align}
\begin{array}{ll}
\label{eq:rel_coh_components}
	H'_{\ua \ub \ug} = - \epsilon_{(\ua \ub} A_{\ug)} , & H'_{\ua \ub \ha} = - \epsilon_{ij} (\Gamma_\ha)_{\ah \bh} \Phi - 
	\epsilon_{ij} \epsilon_{\ah \bh} A_\ha , \\
	H'_{\ua \ha \hb} = \frac{i}{4} (\Sigma_{\ha \hb})_\ah{}^\bh \D_{\bh 
	i} \Phi , &	H'_{\ha \hb \hc} = \frac{3}{8} \D^2_{\ha \hb \hc} \Phi .
\end{array}
\end{align}
The dimension-2 Bianchi identity fixes
\begin{align}
\Phi = \tfrac{i}{24} \D^{\ah i} A_{\ah i}
~~~\mathrm{and}~~~ 
A_\ha = - \tfrac{i}{24} \D^i \Gamma_\ha A_i ,
\end{align}
thus defining all of the components in terms of the spinor potential $A_{\hat \alpha i}$.
The constraints imposed by $\d H' = 0$ on this potential can be presented as
\begin{align}
\label{eq:PhiA1}
	\D_{(\ah (i} A_{\bh) j)} & = 0 , \\[2pt]
\label{eq:PhiA2}
	6 (\Gamma_\ha)_\ah{}^\bh \D_{\bh i} \Phi + 3 (\Sigma_{\ha \hb})_\ah{
	}^\bh \D_{\bh i} A^\hb - (\Sigma_{\ha \hb})_\ah{}^\bh \partial^\hb A_{\bh 
	i} & = 0 , \\[3pt]
\label{eq:PhiA3}
	\D^2_{ij} \Phi & = 0  .
\end{align}

It is illuminating to see precisely how this procedure works. The 1-form $A$ allows the form to ``get off the ground" by giving $H_{\ua \ub \ug}$ a piece to ensure that the lowest Bianchi identity holds even with a scalar superfield sitting inside $H_{\ua \ub \ha}$. However, this is not enough: If we were to continue the analysis with only $A_{\ua}$ and not $A_\ha$ we would find that the final component $H_{\ha \hb \hc}$ vanishes.
Instead, the $A_{\ha}$ component avoids this so that the higher components satisfy the higher Bianchi identities without trivializing.

An interesting feature of this construction is
that, although we are attempting to describe a closed 3-form field-strength, the lower components of this form are not gauge-invariant
under $A_{\ah i} \mapsto A_{\ah i} + \D_{\ah i} \Lambda$ (for some gauge parameter $\Lambda$). 
Nevertheless, the field $\Phi$ is invariant under this transformation so the top two components of $H^\prime$ are invariant (as are the constraints).
This is a generic feature of the relative cohomology construction that comes from solving the closure condition on the form $\beta_{p-1}$ and using its potential $\theta_{p-2}$ in the definition of the closed form $\alpha^\prime_p$.

\section{Field Content in 5D}
\label{sec:5d_content}

The utility of the superforms derived above (and in general) lies in their natural 
accommodation of gauge structure. If we let $A$ be an abelian gauge $(p - 1)$-form, then 
its field-strength $F$ is simply defined as the $p$-form
\begin{equation}
	F = \d A .
\end{equation}
This field-strength is invariant under the gauge transformation $\delta A = \d 
\lambda$ for any $(p-2)$-form $\lambda$, and is itself 
identically closed. 
With the complex laid out in section \ref{sec:5d_forms}, we now 
turn to the field content of the gauge multiplets it defines.

\subsection{The Vector Multiplet (\texorpdfstring{$p = 2$}{p = 2})}
\label{sec:vect_mult}

The theory of a closed, five-dimensional 2-form has at its core a dimension-1 
field-strength $\W$ that satisfies the constraint (\ref{eq:2form_avm_cons}),
identifying it as the field-strength for the five-dimensional vector multiplet of \cite{kuz_lin,how_lin}, as we now review.

Before delving into components and counting degrees of freedom, there are 
two things to note. The first is that by elementary computation,
\begin{equation}
\label{eq:vm_symm}
	\D^{(i}_\ah \D^{j)}_\bh \W = \frac{1}{4} \epsilon_{\ah \bh} \D^{\gh (i} 
	\D_\gh^{j)} \W ~~\Rightarrow~~ \D_\ah^{(i} \D_\bh^j \D_\gh^{k)} \W = 
	0 .
\end{equation}
This will be used later when we look at the degrees of freedom in this multiplet. 
The second thing to note is that by acting on \eqref{eq:2form_avm_cons} with $\D_i^\ah$, 
we obtain for the spinor $\lambda$ in $\W$,
\begin{equation}
\label{eq:vec_mult_dirac}
	\spartial_\ah{}^\bh \lambda_{\bh i} = - \frac{i}{2} \D^2_{ij} \lambda_\ah^j ~\neq~ 0 .
\end{equation}
Thus, this multiplet is off-shell. This may seem curious given that the six-dimensional 
3-form field-strength theory from which this form reduces is on-shell, but note that 
the obstruction to the Dirac equation in \eqref{eq:vec_mult_dirac} is an operator that 
does not exist in six dimensions.

Turning now to the field content, we write the $\theta$-expansion of $\W$ as \cite{kuz_lin}
\begin{equation}
	\W = \phi + i \theta^{\ah i} \lambda_{\ah i} + \frac{i}{2} \theta^{\ah i} \theta_\ah^j X_{
	i j} + i \theta^{\ah i} \theta^\bh_i F_{\ah \bh} + \mathcal{O}(\theta^3) .
\end{equation}
The degrees of freedom in $\W$ are, then,
\begin{equation}
\label{tab:vm_fields}
\begin{tabular}{c|c|c|c|c}
	fields & $\phi$ & $\lambda^\ah_i$ & $X_{ij}$ & $F^{\ah \bh}$ \\
	\hline \hline
	on-shell & ~1~ & 4 & 0 & 3 \\
	\hline
	off-shell & 1 & 8 & 3 & 4 \\
\end{tabular}
\end{equation}
\vspace{0pt}

\noindent since $F_{\ha \hb} = (\Sigma_{\ha \hb})^{\ah \bh} F_{\ah \bh} = - \tfrac{i}{
2} \D^2_{\ha \hb} \W$ and is the field-strength of a dynamical vector due to the dimension-3 Bianchi identity (\ref{eq:VMBI3}). In order to determine the on-shell 
degrees of freedom for the iso-triplet $X_{ij}$, we first need to know whether there 
are any new fields at higher order in $\theta$. To do so, we use the dimension-$\tfrac{
5}{2}$ Bianchi identity \eqref{eq:2form_dim52_bi} and consider what components might live in $\D \D \D \W$. 
To wit, suppose $\D \D \D$ were totally anti-symmetric in spinor indices. If not totally 
symmetric in isospin, the anti-symmetric spinor + anti-symmetric isospin components 
would form partial derivatives. However, if it were totally symmetric in isospin, then 
it would vanish by \eqref{eq:vm_symm}. Therefore the only possible remaining 
source of new components is $\D \D \D$ with at least one symmetric pair of spinor 
indices. But these are exactly the terms that \eqref{eq:2form_dim52_bi} 
rules out. Thus, the fields laid out in \eqref{tab:vm_fields} are the only ones to be 
found and higher components are simply derivatives of the lower ones. Then 
because supersymmetry is required to hold on-shell, $X_{ij}$ is relegated to the 
role of auxiliary field and cannot carry any on-shell degrees of freedom. So with 
this information about the component fields, the action takes the form
\begin{equation}
	\L = \frac{1}{2} \( - \partial^\ha \phi \partial_\ha \phi + i \lambda^i \spartial 
	\lambda_i + \frac{1}{2} X^{ij} X_{ij} - \frac{1}{2} F^{\ha \hb} F_{\ha \hb} 
	- \lambda^i [\phi, \lambda_i] \) .
\end{equation}

\subsection{The Tensor Multiplet (\texorpdfstring{$p = 3$}{p = 3})}
\label{sec:3form_mult}
In section \ref{sec:red_mult} we will discuss the interpretation of the 3-cocycle $H$ of section \ref{sec:5d_3form1}. 
Instead we consider in this section the matter content of the
relative cohomology 3-form $H'$ of section \ref{sec:rc}.
Acting on the constraint (\ref{eq:PhiA2}) with $D^{\hat \alpha}_{(j}$, and using (\ref{eq:PhiA1}) we find that
\begin{align}
\D^2_{\ha ij} \Phi = 0.
\end{align}
This can be combined with the condition (\ref{eq:PhiA3}) to give the superfield constraint\footnote{In the dimensional reduction to $D=4$, this gives the superspace description of the vector-tensor multiplet as it is presented in \cite{Dragon:1997za}. 
} 
\begin{equation}
\label{eq:TMconstraint1}
	\D_\ah^{(i} \D_\bh^{j)} \Phi = 0.
\end{equation}
From this it is straightforward to check that the $\theta$-expansion of $\Phi$,
\begin{equation}
	\Phi = \phi + \theta^\ah_i \chi^i_\ah + \theta^{\ah i} \theta_i^\bh T_{\ah \bh} 
	+ \mathcal{O}(\theta^3),
\end{equation}
stops giving new fields beyond the $\theta^2$-level. Unfortunately, this 
means that the multiplet is an on-shell tensor multiplet with the degrees 
of freedom
\begin{equation}
\label{tab:tm_fields}
\begin{tabular}{c|c|c|c}
	fields & $\phi$ & $\chi^\ah_i$ & $T^{\ah \bh}$ \\
	\hline \hline
	on-shell & ~1~ & 4 & 3
\end{tabular}
\end{equation}
where $T_{\ah \bh} = \frac12(\Sigma^{\ha \hb})_{\ah \bh} T_{\ha \hb}$ 
is dual to the 3-form field-strength $F_{\ha \hb \hc}$ of a 2-form gauge field.
(Alternatively, we may observe that equation (\ref{eq:TMconstraint1}) is the vector multiplet constraint (\ref{eq:2form_dim2_acons}) and the condition (\ref{eq:PhiA3}) is its equation of motion \cite{kuz_lin}.)
These component fields imply that an action takes the form
\begin{equation}
	\L = \frac{1}{2} \( - \partial^\ha \phi\, \partial_\ha \phi + i \chi^i \spartial 
	\chi_i  + \frac{1}{6} F^{\ha \hb \hc} F_{\ha \hb \hc} 
	\) .
\end{equation}

\subsection{The Linear Multiplet (\texorpdfstring{$p = 4$}{p = 4})}
\label{S:Linear}

The supermultiplet content described by a closed, five-dimensional 4-form 
is contained inside a superfield $G_{ij}$ subject to the analyticity constraint
\begin{equation}
\label{eq:4form_mult_cons}
	\D_{\ah (i} G_{jk)} = 0 .
\end{equation}
This is the five-dimensional, $\N = 1$ linear multiplet, the four-dimensional $\N = 2$ version
of which was discovered in \cite{sohnius_lm}.\footnote{A five-dimensional formulation is given in \cite{bkn} but they do not examine the field content before reducing to a centrally-extended 4D, $\N = 2$ superspace.} 
The $\theta$-expansion 
is
\begin{equation}	
G_{ij} = \varphi_{ij} + 2 \theta_{(i} \psi_{j)} + 2i \theta_i 
\Gamma^\ha \theta_j V_\ha + \theta_i \theta_j M +  \text{derivatives},
\end{equation}
where $\varphi_{ij}$ is an iso-triplet of scalars, $\psi_\ah^i$ is a doublet of 
Weyl fermions, $V_\ha$ is a vector field-strength, and $M$ is a real auxiliary 
scalar. Additionally, the constraint \eqref{eq:4form_mult_cons} requires that 
$\partial_\ha V^\ha = 0$. This condition can be solved as
\begin{equation}
	V^\ha = \epsilon^{\ha \hb \hc \hd \he} \partial_\hb E_{\hc \hd \he}
\end{equation}
for a gauge 3-form $E$. The degrees of freedom carried by these fields are

\begin{equation}
\label{tab:4form_fields}
\begin{tabular}{c|c|c|c|c}
	fields & $\varphi_{ij}$ & $\psi_\ah^i$ & $E^{\ha \hb \hc}$ & $M$ \\
	\hline \hline
	on-shell & 3 & 4 & 1 & 0 \\
	\hline
	off-shell & 3 & 8 & 4 & 1 \\
\end{tabular}
\end{equation}
\vspace{0pt}

\noindent and so the supermultiplet is off-shell. Finally, the action for this multiplet is
\begin{equation}
	\L = \frac{1}{2} \( \frac{1}{2} \partial_\ha \varphi^{ij} \partial^\ha \varphi_{ij} 
	- V^\ha V_\ha + i \psi^i \spartial \psi_i + M^2 \).
\end{equation}

The component field content of this section also indicates a relation to the results
of \cite{LinN2,bkn}.  When one reduces the component field content of the 3-form
$E_{\hc \hd \he}$ to four dimensions, one obtains 2-form gauge field $E_{c d 5}$
 and a four dimensional gauge 3-form $E_{c d e}$.  Then the $\N = 1$
 supermultiplet content is seen to be $\left( \varphi_{22} ,  \psi_2,  E_{c d 5}  \right)$
 and $\left( \varphi_{11} ,  \varphi_{12},  \psi_1 ,  E_{c d e} ,  M \right)$.  The
 first of these is a $\N = 1$ tensor multiplet and the second is a variant formulation
 of a  $\N = 1$ chiral supermultiplet \cite{var}.  The latter of these contains one 0-form auxiliary field
 $M$ and a 3-form auxiliary field $E_{c d e} $.

\subsection{Reducible Multiplets}
\label{sec:red_mult}

We have found that the procedure of obstructing the Bianchi identities of an irreducible supersymmetric multiplet describing a $p$-form generally fails to give an irreducible multiplet describing a $(p+1)$-form. To distinguish these cases, we will refer to the elements of the super-de Rham complex as constructed here as ``$p$-cocycles''. When these have an interpretation as an irreducible supermulitplet containing a closed bosonic $p$-form, we will call them closed (super-)$p$-forms.

Examples of cocycles that are not closed forms were found in section \ref{sec:5d_2form_alt} for $p=2$ and in section \ref{sec:5d_3form1} for $p=3$. In the first case, there were two 2-cocycles, one of which is a closed 2-form. In the latter, however, there was no de Rham 3-cocycle that could be interpreted as a 3-form. (For this, we had to pass to the 3-cocycle of the relative cohomology of section \ref{sec:rc}.) 
From the four-dimensional perspective, this is a new phenomenon: At least in the case of 4D, $N=1$, every $p$-cocycle is a closed $p$-form.

What, then, is the interpretation of such cocycles? A clue is to be found by scrutinizing the constraints on the field-strengths of cocycles that are closed forms. In very low degree, the $p$-cocycles are guaranteed to be forms since we can always start with a scalar superfield and take its derivative to get an exact 1-form. Similarly, in high degree, specifically co-dimension 1, the $(D-1)$-cocycle has the interpretation of a closed $(D-1)$-form because 
its analyticity implies that it contains a conserved vector field-strength,
as described in section \ref{S:Linear}. When $D \leq 4$, the 2-form field-strength (guaranteed to exists as the Maxwell field-strength), sits directly beneath the $D-1=3$-form field-strength. However, when $D > 4$ a gap opens up between $p=2$ and $p=D-1$ and it is in this gap that we find a cocycle that is not guaranteed to have an interpretation as a closed form. In fact, both of the non-form cocycles we have found are naturally associated to the co-dimension-1 form of sections \ref{sec:5d_45forms} and \ref{S:Linear}, as we can see from the progression of constraints
\begin{align}
	\Pi_{\ha \hb \ah}^{~ \hc \hd \bh} D_{\bh (i} C_{\hc \hd jk)} 
		\stackrel{(\ref{eq:2form_cons2})}= 0
~,~~	
	\Pi_{\ha \ah}^{~\hb \bh} D_{\bh (i} H_{\hb jk)}
		\stackrel{(\ref{eq:3form_cons2})}= 0
~,~\mathrm{and}~~
	\Pi_\ah^{~\bh} D_{\bh (i} G_{jk)}
	\stackrel{(\ref{eq:4form_cons})}= 0
~,
\end{align}
where the $\Pi$s are the projectors (cf. eqs. \ref{eq:pi_stls}, \ref{eq:pi_gtls}, and taking $\Pi_\ah^{~\bh}:= \delta_\ah^{\bh}$) onto the anti-symmetric tensor, vector, and scalar representations, respectively.

Alternatively, it is not the expectation that there be a closed form interpretation of the cocycle that fails insomuch as it is that the cocycle may be required to be a {\em composite} closed form. Consider, for example, the 
$2$-cocycle $A\wedge A^\prime$ constructed by wedging two different 1-forms. The lowest component of this product generally contains both the 2-form part $\sim A^{\hat \alpha i}A^\prime_{\hat \alpha i}$ from section \ref{sec:5d_2form} and the $2^\prime$-cocycle part $\sim A_{(\hat \alpha (i}A^\prime_{\hat \beta) j)}$ from section \ref{sec:5d_2form_alt}. Therefore, the existence of the $2^\prime$-cocycle is required by the fact that differential forms form a differential graded algebra with respect to the $\wedge$-product.

We conclude with a related observation for which we do not yet have a complete explanation: The 3-cocycle $H$ of section \ref{sec:5d_3form1} satisfies the constraints of one of the five-dimensional, $N=1$ conformal supergravity torsions worked out in reference \cite{Kuzenko:2008wr}. Specifically, this superspace contains a dimension-1 torsion $C_{\hat a  ij}$ constrained by the dimension-$\frac32$ Bianchi identities to satisfy equation (\ref{eq:3form_cons1}). Under local superconformal transformations, $\delta C_{\hat a  ij} = \sigma C_{\hat a  ij} -i D^2_{\hat a  ij} \sigma$. The first term is the transformation of a superconformal primary field of weight 1 and the inhomogeneous term indicates that $C$ is a connection for local superconformal transformations. In this sense, the cocycle $H_{\hat a  ij} \sim D^2_{\hat a  ij} \sigma$ describes the gauge parameters of local superconformal transformations in five-dimensional superspace.\footnote{The analogous thing happens in six dimensions in terms of the 4-cocycle.
} 

\section{Conclusions}
In this article we have constructed the super-de Rham complex in five-dimensional, $\N = 1$ superspace and related it to the complex of six-dimensional, $\N = (1, 0)$ superspace \textit{via} dimensional reduction. This turned out to be only one part of the reduced complex, with the remaining part serving as an additional source of closed superforms arising from the relative cohomology of the two superspaces. A surprising feature of the five-dimensional complex is that the 3-form field-strength $H$ does not describe an irreducible supermultiplet arising from the supersymmetrization of a closed bosonic $3$-form. Instead, the ``missing'' tensor multiplet arises from the relative cohomology construction of section \ref{sec:rc}.

We concluded our excursion in 5D by investigating the field content described by the $p$-form field-strengths for $p = 2,\, 3,\, 4$ which were, respectively, an off-shell vector multiplet, an on-shell tensor multiplet, and an off-shell linear multiplet (with gauge 3-form). The 4-form field-strength also automatically solved
a problem left open from the work of \cite{LinN2}; namely, by dimensional
reduction of the results in section \ref{S:Linear} we have found the 4D, $\N = 2$ supermultiplet containing a component level 3-form gauge field.

In this paper we have taken steps to fill in our understanding of eight-supercharge superspaces as we bracket our work with the extensive literature on $\R^{4 \vert 8}$ and the six-dimensional complex of \cite{6dforms}. However, we have also uncovered questions that should extend beyond specific superspaces and hint towards a more universal understanding of superforms. In the associated works \cite{lin_ran, randall} we study the problem noted in section \ref{sec:5d_3form1} of determining how constraints fit together inside Bianchi identities generically and examine the dimensional reduction for embedded superspaces $\R^{D - 1 \vert n}\hookrightarrow \R^{D \vert n}$.

Finally, we note that this work has introduced new curiosities about how superforms may be used to discover superfield formulations of gauge supermultiplets. In higher dimensions it appears to now be an open question as to how certain gauge theories can be constructed. The example we encountered in five dimensions is that the superform description of an off-shell tensor multiplet in ordinary 5D, $\N = 1$ superspace ({\it i.e.} without central charge and/or harmonics) remains unknown. If we try to obtain such a superform by either of the dimensional reduction paths laid out in section \ref{sec:dim_red}, we obtain a multiplet of superconformal gauge parameters or an on-shell tensor multiplet. If we instead start in 4D, $\N = 2$ superspace with the vector-tensor multiplet, this lifts to five dimensions by becoming the on-shell tensor multiplet.

There are also other extensions to flat superspace that may be considered; 4D, $\N = 2$ centrally-extended superspaces have been considered in \cite{bkn, kuz_nov, novak} and have a close relationship with 5D, $\N = 1$ given that the central charge can be considered a $\partial_5$ term. Curved superspaces are another area of interest as we consider how such spaces fit into the general discussion of superform constraints and dimensional reduction. Work on these topics is underway at the present time as we continue our march towards understanding the geometry of superspace and its relationship to the structure of gauge theories in arbitrary dimension with any number of superysmmetries.

\section{Acknowledgments}

This work was partially supported by the National Science Foundation 
grants PHY-0652983 and PHY-0354401 and the University of Maryland Center for String \& Particle Theory. 
SR was also supported by the Maryland Summer Scholars program and the Davis Foundation and participated in the 2013 and 2014 Student Summer Theoretical Physics Research Sessions. 
WDL3 thanks the Simons Center for Geometry and Physics for hospitality during the XII Simons Workshop. 


\appendix

\section{Five-dimensional, $N=1$ Superspace}
\label{sec:susy_math}
\setcounter{equation}{0}

Our five-dimensional notation and conventions were first given in \cite{kuz_lin} and are designed to reduce to those of \cite{Buchbinder:1998qv} in 4D. Using the ``mostly-plus" flat metric $\eta_{\ha \hb}$, for $\ha, \hb \in \{ 0, 1, 2, 3; 5\}$, our $\Gamma$-matrices $\Gamma_\ha = (\Gamma_a, \Gamma_5)$, with $a \in \{0, 1, 2, 3\}$, are chosen to satisfy the algebra
\begin{equation}
\label{eq:5d_clifford}
	\{ \Gamma_\ha, \Gamma_\hb \} = - 2 \eta_{\ha \hb} \mathbf{1}.
\end{equation}
In order to completely span the space of $4 \times 4$ matrices we introduce the symmetric matrices $\Sigma_{\ha \hb} := - \frac{1}{4} [\Gamma_\ha, \Gamma_\hb]$ to complement the anti-symmetric spinor metric $\epsilon_{\ah \bh}$ and anti-symmetric, traceless $\Gamma$-matrices.

We also have the useful identities for $A_{ij} = A_{[ij]}$:
\begin{equation}
	A_{ij} = \tfrac{1}{2} \epsilon_{ij} A^k{}_k \qquad \text{and} \qquad A^{ij} = - \tfrac{
	1}{2} \epsilon^{ij} A^k{}_k,
\end{equation}
where $\epsilon_{ij}$ is the isospinor metric. The algebra of 5D, $\N = 1$ superspace 
is then
\begin{equation}
\label{eq:5dN1_alg}
	\{ \D_\ah^i, \D_\bh^j \} = - 2 i \epsilon^{ij} \spartial_{\ah \bh},
\end{equation}
where, for reference, the $\D$s are explicitly defined as
\begin{equation}
	\D_{\ah i} := \partial_{\ah i} - i \spartial_{\ah \bh} \theta^\bh_i.
\end{equation}
The irreducible $\D^2$ operators in five dimensions are normalized as follows:
\begin{align}
	\D^2_{ij} := \tfrac{1}{2} \D_{(i} \D_{j)}, \quad \D^2_{\ha ij} := \tfrac{1}{2} \D_{(i} \Gamma_\ha \D_{j)}, \quad \text{and} \quad \D^2_{\ha \hb} := \tfrac{1}{2} \D^i \Sigma_{\ha \hb} \D_i.
\end{align}
Note that here we use the contraction convention $\psi^{\ah i} \chi_{\ah i} = \psi \chi$. With these operators, we can expand a generic $\D \D$ object as
\begin{align}
\label{eq:dd_exp}
	\D_{\ah i} \D_{\bh j} & = i \epsilon_{ij} \spartial_{\ah \bh} - \tfrac{1}{2} \epsilon_{
	ij} (\Sigma^{\ha \hb})_{\ah \bh} \D^2_{\ha \hb} + \tfrac{1}{2} \epsilon_{\ah \bh} 
	\D^2_{ij} + \tfrac{1}{2} (\Gamma^\ha)_{\ah \bh} \D^2_{\ha ij}.
\end{align}
We also define the shorthand
\begin{equation}
	\D^2_{\ha \hb \hc} := - \tfrac{1}{12} \epsilon_{\ha \hb \hc}{}^{\hd \he} \D^2_{\hd 
	\he}
\end{equation}
so that
\begin{equation}
	\epsilon_{\ha \hb}{}^{\hc \hd \he} \D^2_{\hc \hd \he} = \D^2_{\ha \hb} .
\end{equation}
Straightforward $\D$-pushing with the algebra \eqref{eq:5dN1_alg} yields the 
following commutators
\begin{align}
	[\D^{2 ij}, \D^2_{\ha ij}] & = 12i \partial^\hb \D^2_{\ha \hb} , \\
	[\D^{2 ij}_\ha , \D^2_{\hb ij}] & = 72i \partial^\hc \D^2_{\ha \hb \hc} , \\
	[\D^2_{ij},\, \D^2_{\ha \hb}] & = - 4i \partial_{[\ha} \D^2_{\hb] ij}
\end{align}
which are useful in the calculations of section \ref{sec:5d_forms}.

It will also be helpful to note some elementary facts about $\D^3$ operators. As shown by Koller \cite{koller}, in six dimensions there are only two linearly independent $\D^3$s; namely, $\D^3_{\alpha ijk}$ and $\tilde{\D}^3_{a \alpha i}$. In five dimensions the vector component of $\tilde{\D}^3$ splits, and so we have three:
\begin{align}
	\tilde{\D}^3_{\ah i} := \{ \D_\ah^j, \D^2_{ij} \} 
~,~	
	\tilde{\D}^3_{\ha \ah i}  := \{ \D_\ah^j, \D^2_{\ha ij} \} 
~,~	
	\D^3_{\ah ijk} := \{ \D_{\ah (i}, \D^2_{jk)} \} = 2 \D_{\ah (i} \D^2_{jk)}.
\end{align}
These definitions lead to the relations
\begin{align}
	\{ \D_{\ah i}, \D^2_{jk} \} & = \D^3_{\ah ijk} + \frac{2}{3} \epsilon_{i (j} \tilde{\D}^3_{k) \ah} , \cr
	\{ \D_{\ah i}, \D^2_{\ha jk} \} & = - (\Gamma_\ha)_\ah{}^\bh \D^3_{\bh ijk} + \frac{2}{3} \epsilon_{i (j} \tilde{\D}^3_{k) \ah \ha} , \cr
	\{ \D_{\ah i}, \D^2_{\ha \hb} \} & = \frac{2}{3} (\Gamma_{[\ha})_\ah{}^\bh \tilde{\D}^3_{\hb] \bh i} + \frac{2}{3} (\Sigma_{\ha \hb})_\ah{}^\bh \tilde{\D}^3_{\bh i} ,
\end{align}
where we've used the fact that\footnote{This is consistent with the 6D condition $(\tilde{\gamma}^a)^{\alpha \beta} \tilde{\D}^3_{a \beta i} = 0$.}
\begin{equation}
	(\Gamma^\ha)_\ah{}^\bh \tilde{\D}^3_{\ha \bh i} = - \tilde{\D}^3_{\ah i} .
\end{equation}
We can now expand a generic $\D \D \D$ object by decomposing any two $\D$s using \eqref{eq:dd_exp} and then writing the $\D \D^2$ terms as $[\D, \D^2] + \{\D, \D^2\}$.

Finally, we note the following $\Gamma$-matrix identities that follow directly from \eqref{eq:5d_clifford} as worked out in \cite{kuz_tm}: the 
completeness relation
\begin{equation}
\label{eq:5d_completeness}
	\epsilon_{\ah \bh \gh \dh} = \frac{1}{2} (\Gamma^\ha)_{\ah \bh} (\Gamma_\ha
	)_{\gh \dh} + \tfrac{1}{2} \epsilon_{\ah \bh} \epsilon_{\gh \dh} ,
\end{equation}
the trace identities
\begin{equation}
	\operatorname{tr} \Gamma^\ha \Gamma^\hb = - 4 \eta^{\ha \hb} \qquad \text{
	and} \qquad \operatorname{tr} \Sigma^{\ha \hb} \Sigma_{\hc \hd} = - 2\delta^{
	[\ha}_{[\hc} \delta^{\hb]}_{\hd]} ,
\end{equation}
and the expansions
\begin{align}
	(\Gamma^\ha)_\ah{}^\gh (\Gamma^\hb)_\gh{}^\bh & = - \eta^{\ha \hb} 
	\delta_\ah^\bh - 2 (\Sigma^{\ha \hb})_\ah{}^\bh , \cr
	(\Gamma^\ha)_\ah{}^\gh (\Sigma^{\hb \hc})_\gh{}^\bh & = - \frac{1}{2} 
	\epsilon^{\ha \hb \hc \hd \he} (\Sigma_{\hd \he})_\ah{}^\bh + \eta^{\ha [\hb} 
	(\Gamma^{\hc]})_\ah{}^\bh .
\end{align}


\end{document}